\documentclass[twocolumn, superscriptaddress,
 amsmath,amssymb,
 aps,prb,
]{revtex4-2}

\usepackage{graphicx}
\usepackage{dcolumn}
\usepackage{bm}
\usepackage{physics}
\usepackage{here}
\usepackage{xcolor}
\usepackage{braket}
\usepackage[
 colorlinks=true,
 allcolors=blue,
]{hyperref}

\begin{document}
\title{Chern insulators and topological flat bands in cavity-embedded kagome systems}

\author{Hikaru Goto}
\author{Ryo Okugawa}
\author{Takami Tohyama}
\affiliation{Department of Applied Physics, Tokyo University of Science, Tokyo 125-8585, Japan}
 
\date{\today}

\begin{abstract}
We investigate topological band structures of a kagome system coupled to a circularly polarized cavity mode,
using a model based on a muffin-tin potential and quantum light-matter interaction.
We show that Chern insulating phases emerge in the cavity-embedded kagome system due to the light-matter interaction that breaks time-reversal symmetry.
We also find that a nearly flat band can be topologically nontrivial with a nonzero Chern number.
By varying the light-matter interaction,
we also reveal that topological phase transitions occur between different Chern insulating phases in the ultrastrong coupling regime.
The phase transitions change the sign of the Chern number,
switching the direction of the edge current.
We demonstrate the existence of topological edge modes in the cavity-embedded kagome Chern insulators
by constructing a low-energy effective tight-binding model.
\end{abstract}

\maketitle

\section{INTRODUCTION}
The manipulation of quantum phases via light-matter coupling has attracted significant attention in condensed matter physics~\cite{Kockum2019_cavity_review, FornDiaz2019_cavity_review, Torre2021_cavity_review, Schlawin2022_cavity_review, Bloch2022_cavity_corre, Lu2025_cavity_review, Baydin2025_cavity_review, Mueller2025_cavity_review}.
Optical cavities provide a powerful platform for tuning light-matter coupling between electrons and photons. 
Recent experiments have achieved strong light-matter interactions in cavities where the coupling strength exceeds a fraction of the bare cavity frequency,
which is known as the ultrastrong coupling (USC) regime~\cite{Maissen2014_cavity_usc, 
Keller17, Paravicini2019_cavity_usc, Appugliese2022_breakdown_topo, Andberger2024, Enkner2025_cavity_qhe}.
Because electronic states can be drastically modified in the USC regime,
cavities provide a route for studying unconventional physics.
Along with such experimental developments, 
recent theoretical studies have investigated various many-body phenomena in cavity quantum electrodynamics (QED), 
such as ferroelectricity~\cite{Ashida2020_cavity_ferroele, Curtis2023_cavity_ferroele}, superconductivity~\cite{Sentef2018_cavity_sc, Schlawin2019_cavity_sc, Kozin2025_cavity_sc}, quantum spin liquids~\cite{Chiocchetta2021_QSL, Masuki2024_moire, wei2025_cavity_corre}, and Kondo effect~\cite{Kuo2023, Mochida2024_Kondo_corre}.

Furthermore, control of topological phases with the cavity has been explored extensively~\cite{Kibis2011_cavity_graphene, Wang2019_cavity_graphene,  Li2022_cavity_topo, Dmytruki2022_cavity_topo, Appugliese2022_breakdown_topo, Masuki2023_berryphase, Zuzhang2023_cavity, Nguyen2023_epChern, Rokaj2023_cavity_qhe, Jiang2024_cavity, Bacciconi2024_cavity, Shaffer2024_cavity_topo, Dag2024_cavity_topo, PèrezGonzàlez2024_topo, Nguyen2024_cavity_topo, Yang2025_EHM_topo, Bacciconi2025_cavity, ghorashi2025_cavity_topo_moire, Enkner2025_cavity_qhe, shin2025_cavity, Cardoso2026_cavity, Yang2026_cavity}.
For example, graphene has been predicted to host a Chern insulating phase under a circularly polarized cavity field because Dirac points are gapped out by the light-matter interaction~\cite{Masuki2023_berryphase, Yang2025_EHM_topo, Kibis2011_cavity_graphene, Wang2019_cavity_graphene, Li2022_cavity_topo, Dag2024_cavity_topo}.
These topological phase transitions do not necessarily require strong light-matter interaction
if time-reversal symmetry is broken \cite{Haldane1988}.
A topological phase that appears only in the USC regime has yet to be found in graphene.
So far, topological phases in the USC regime have remained elusive.

Kagome materials have been regarded as a promising platform for Chern insulating phases
and topological flat-band systems
because their band structure possesses Dirac points and a nearly flat band with band degeneracy \cite{Yin2022_kagome, Wang2024_kagome_review, Li2025_kagome_review, Wang2025_kagome_review}.
Since time-reversal symmetry breaking leads to a mass gap  \cite{Ohgushi2000_kagomeQAH},
magnetic kagome systems are expected to host topological bands characterized by a Chern number ~\cite{Ye2018_kagome, Yin2020_kagome, Li2021_kagome, Ma2021_kagome}.
In kagome systems without time-reversal symmetry,
flat bands can be topologically nontrivial with a nonzero Chern number
\cite{Ohgushi2000_kagomeQAH, Tang2011, Rhim2019, Okamoto2022, Baidya2019_kagome}.
In view of topological band engineering,
time-reversal symmetry breaking is essential for realizing such topological phases.
Given the unique band structure of kagome systems,
modifying their bands inside an optical cavity can open up the possibility of realizing various topological band structures from the weak coupling to the USC regime.

In this paper, we investigate topological phases of the kagome system in a cavity with circularly polarized light.
We show its band structures and topological phases from weak to strong coupling regimes.
By introducing the cavity field, we show that the kagome system becomes a Chern insulator because the band degeneracy is lifted.
We also demonstrate that the light-matter interaction produces a nearly flat band with a nonzero Chern number.
Moreover, we reveal that in the USC regime, various Chern insulating phases appear in the kagome system, unlike a honeycomb system that shows only one Chern insulating phase.
We also show topological gapless edge states in the kagome Chern insulators.

The rest of this paper is organized as follows.
In Sec.~\ref{sec:model}, we introduce a model for kagome and honeycomb systems embedded in a chiral cavity.
We analyze band structures and topological phases of the model over a broad range of coupling strengths of the light-matter interaction in Sec.~\ref{sec:bulk}.
To study topological edge states in the Chern insulating phases,
we construct a low-energy tight-binding model and demonstrate the presence of the edge states in Sec.~\ref{sec:edge}.
Finally, a summary is given in Sec.~\ref{sec:conc}.

\section{MODEL}
\label{sec:model}

We begin with a cavity QED Hamiltonian in the Coulomb gauge, which is described as~\cite{Masuki2023_berryphase}
\begin{align}
    \hat{H}_{\mathrm{C}}=  \frac{(\hat{\bm{p}} + e \hat{\bm{A}})^2}{2m} + V(\bm{r}) 
    +\hbar \omega_{\mathrm{c}} \left( \hat{a}^{\dagger} \hat{a}  + \frac{1}{2} \right), 
    \label{eq:QED_Hamil_Coulomb}
\end{align}
where $m$ and $-e$ are an effective mass and the charge of an electron, respectively, and $V(\bm{r})$ is an arbitrary periodic potential.
In this paper, we neglect both direct and cavity-mediated electron-electron interactions,
and calculate band structures
\footnote{
To calculate band structures in momentum space, we apply the transformation $\hat{H}_{\mathrm{C}}(\bm{k}) = e^{-i\bm{k}\cdot\bm{r}} \hat{H}_{\mathrm{C}} e^{i\bm{k}\cdot\bm{r}}$.
We then diagonalize $\hat{H}_{\mathrm{C}}(\bm{k})$ by expanding the eigenstates in the basis of the product state of a plane wave state and a photon number eigenstate $|\bm{\mathcal{K}} \otimes n\rangle$,
where $\bm{\mathcal{K}}$ is a reciprocal lattice vector and $n$ is the photon number,
and obtain the corresponding energy eigenvalues
}.
Here, $\hat{\bm{A}} = A_0 (\bm{\varepsilon} \hat{a} + \bm{\varepsilon}^* \hat{a}^{\dagger})$ is the quantized vector potential with the mode amplitude $A_0$, where $\hat{{a}}\ (\hat{{a}}^{\dagger})$ is the annihilation (creation) operator of a photon with a cavity mode $\omega_{\mathrm{c}}$ and $\bm{\varepsilon} = (1, -i)^{\mathrm{T}} / \sqrt{2}$ is the circular polarization vector that breaks time-reversal symmetry.
Such a circularly polarized cavity mode can be realized in chiral cavities
\cite{Hubener2021}.
The coupling strength of the light-matter interaction is quantified by $g = eA_0 \sqrt{\omega_{\mathrm{c}} / (m \hbar)}$
\cite{Ashida2021_ADframe, Masuki2023_berryphase}.
We vary the coupling strength by changing the vector potential.
If $g/\omega_{\mathrm{c}} \gtrsim 0.1$, this parameter range is typically called the USC regime.

\begin{figure}
    \centering
    \includegraphics[width=0.8\linewidth]{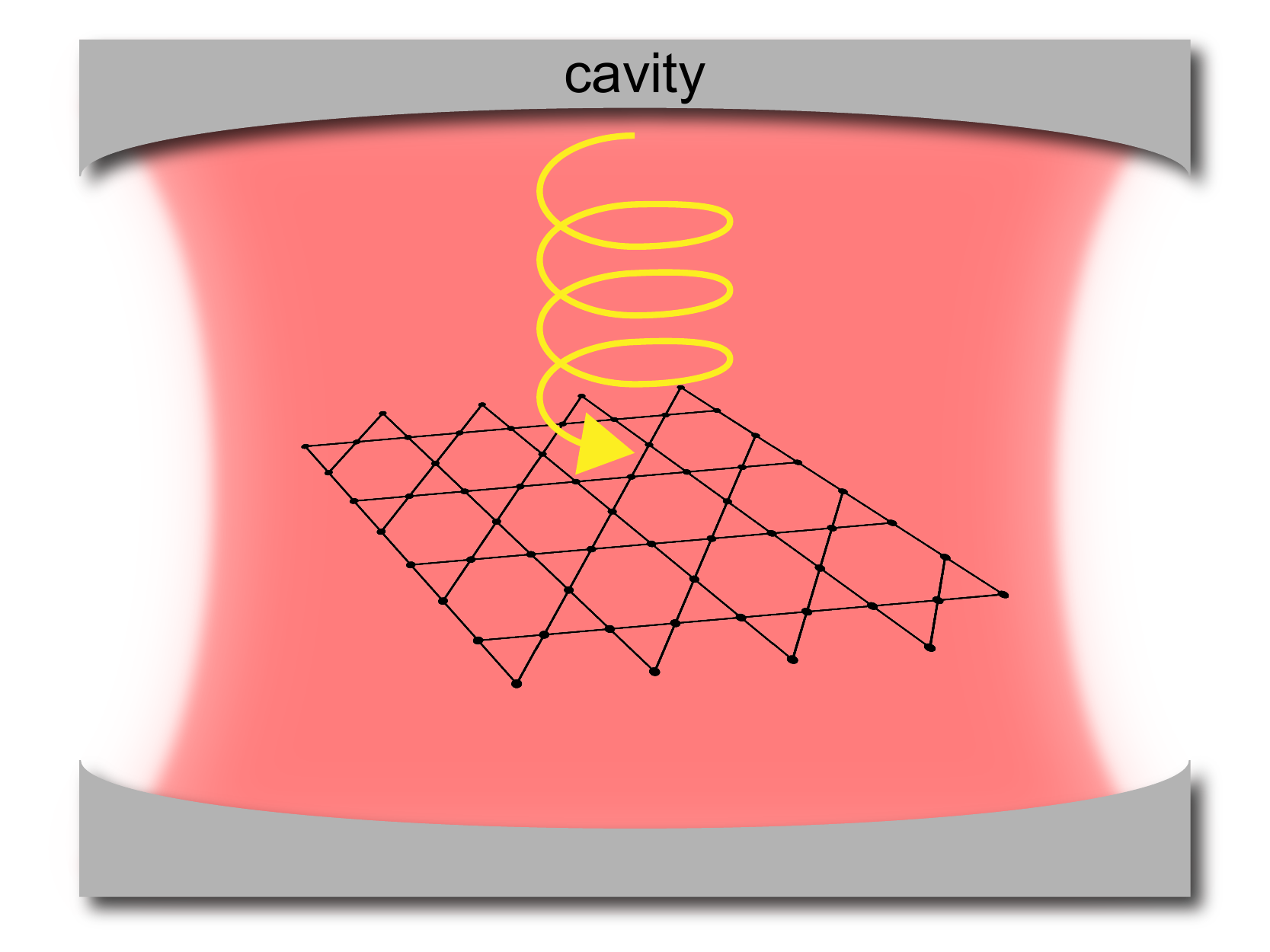}
    \caption{
    Schematic drawing of a kagome material embedded in a chiral cavity.
    The arrow represents a circularly polarized cavity mode.}
    \label{fig:schematic}
\end{figure}

\begin{figure}[t]
    \centering
    \includegraphics[width=1.0\linewidth]{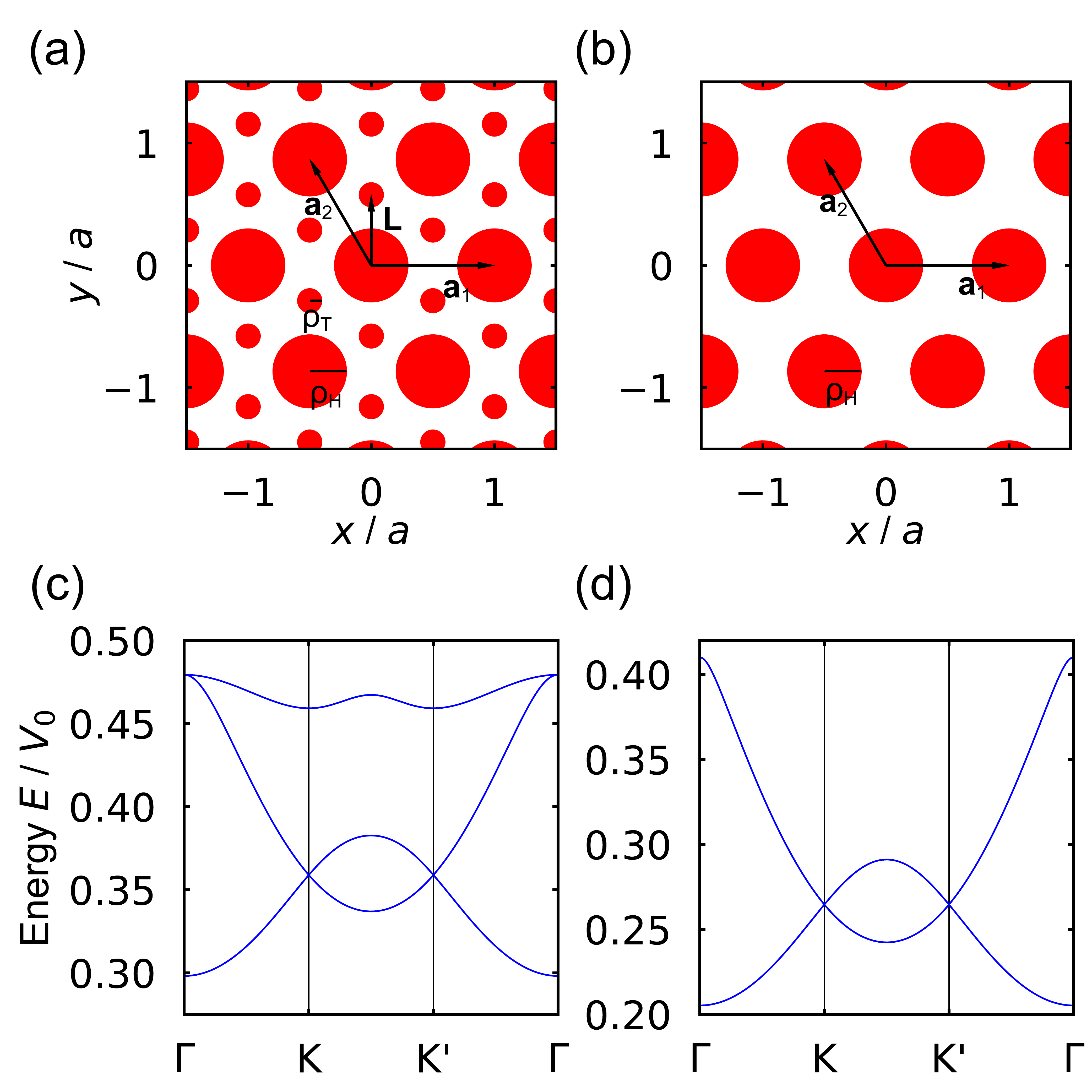}
    \caption{(a) and (b) Periodic potential for kagome and honeycomb lattices, respectively.
    In (a), the center of the small circles corresponds to that of the triangular unit cells in the kagome lattice.
    In (b), the small circles are absent because the potential terms with radius $\rho_{\mathrm{T}}$ vanish.
    (c) and (d) Energy bands for $\alpha=1$ and $\alpha =0$, respectively.
    }
    \label{fig:periodic_bareband}
\end{figure}

We study kagome and honeycomb systems coupled to a circularly polarized mode in a chiral cavity,
as illustrated in Fig.~\ref{fig:schematic}.
Such systems can be described by the following muffin-tin potential:
\begin{align}
\begin{split}
V(\bm{r}) = V_0 \sum_{\bm{R}} &\Big[ \theta(\rho_{\mathrm{H}} - \lvert \bm{r} - \bm{R} \rvert)  \\
    &+\alpha \sum _{s=\pm 1}\theta(\rho_{\mathrm{T}} - \lvert \bm{r} - \bm{R} +s \bm{L} \rvert) \Big] ,
\end{split}
  \label{eq:muffin-tin_potential}
\end{align}
where $\theta$ denotes the Heaviside step function, and
$\bm{R}$ is a lattice vector given by the primitive vectors $\bm{a}_1 = (a,\ 0)$ and $\bm{a}_2 = (-a/2,\ \sqrt{3}a /2)$ with lattice constant $a$, and $\bm{L} = ( 0, a / \sqrt{3} )$.
As shown in Figs.~\ref{fig:periodic_bareband}(a) and \ref{fig:periodic_bareband}(b), the potential $V(\bm{r})$ consists of barriers with height $V_0>0$ and radius $\rho_{\mathrm{H}}$ located at the lattice sites $\bm{R}$, and additional barriers with height $\alpha V_0$ and radius $\rho_{\mathrm{T}}$ centered at $\bm{R} \pm\bm{L}$.
Here, $\alpha $ is a real dimensionless parameter, 
and the potential yields a kagome lattice for $\alpha =1$, and a honeycomb lattice for $\alpha =0$.
Without vector potential, the band structures of the system with $\alpha =1$ have been studied as a kagome model \cite{Xu2024_kagome_Muffintin}.
We note that tunable honeycomb and kagome band structures have been experimentally realized in GaAs-based two-dimensional electron gases (2DEGs) using artificially patterned potentials \cite{Singha11Science,Wang2018,wang2024tuning}.

To confirm whether the potential in Eq.~(\ref{eq:muffin-tin_potential}) imitates typical kagome and honeycomb lattices, we compute bulk energy bands in the absence of the light-matter interaction.
We set $m=90\hbar^2 /(a^2 V_0), \hbar\omega_{\mathrm{c}}=0.2V_0, \rho_{\mathrm{H}}=0.3a$, and $\rho_{\mathrm{T}}=\rho_{\mathrm{H}}/3$,
and use these parameters throughout this paper.
For example, if $V_0=10.2~\mathrm{meV}$ and $a=100~\mathrm{nm}$ are chosen,
we obtain $m=0.067 m_{\mathrm{e}}$ with the bare electron mass $m_{\mathrm{e}}$,
which is in line with the effective mass of GaAs-based 2DEGs with patterned potential \cite{wang2024tuning}.
Figure \ref{fig:periodic_bareband} (c) shows the three lowest energy bands for $\alpha =1$.
We can find Dirac points at the $\mathrm{K}=2\pi/a(1/3, 1/\sqrt{3})$ and $\mathrm{K}' =2\pi/a(2/3, 0)$ points and
a nearly flat band with band degeneracy at the $\Gamma=(0,0)$ point.
This band structure is consistent with the previous study \cite{Xu2024_kagome_Muffintin},
and is similar to that of typical kagome systems \cite{Yin2022_kagome, Wang2024_kagome_review, Li2025_kagome_review, Wang2025_kagome_review}.
We also show bulk bands for $\alpha =0$ in Fig.~\ref{fig:periodic_bareband} (d).
In this case, the nearly flat band is shifted to higher energy.
In contrast, the Dirac points remain at the $\mathrm{K}$ and $\mathrm{K}'$ points in the two lowest bands,
which can be found in honeycomb lattice models such as graphene \cite{Castro09}.
Hence, we call the systems with $\alpha =1$ and $\alpha =0$ kagome and honeycomb systems, respectively.

\section{Chern insulating phases and topological flat bands induced by the cavity field}
\label{sec:bulk}

\begin{figure*}[t]
    \centering
    \includegraphics[width=1.0\linewidth]{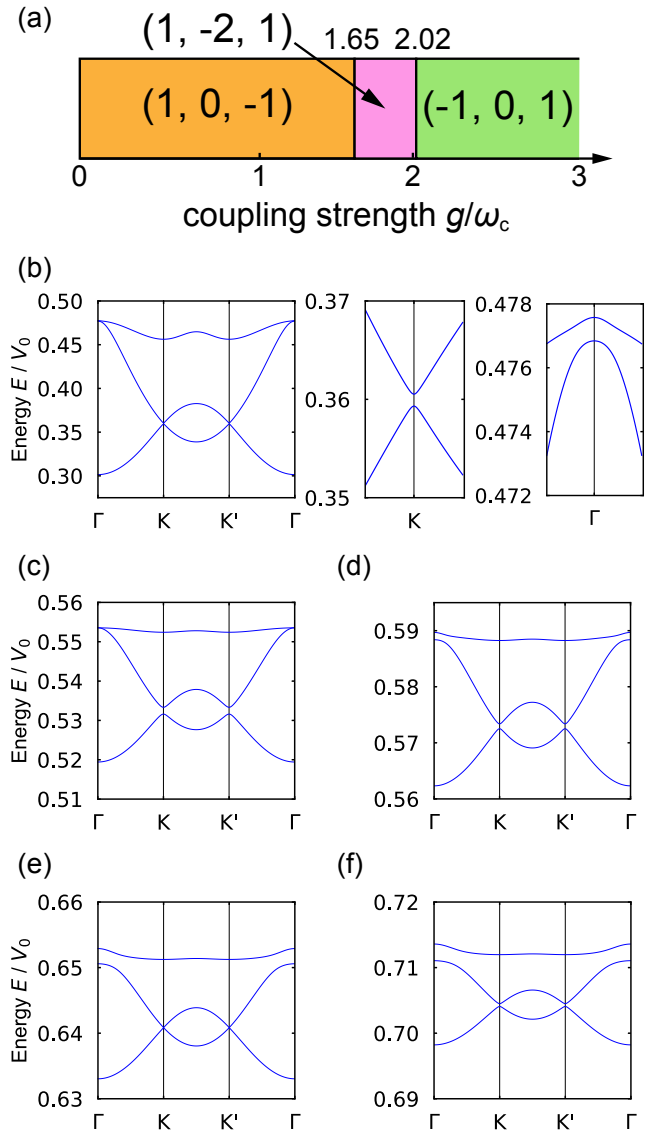}
    \caption{(a) Chern numbers of the kagome system with the circularly polarized cavity field.
    Each band is topologically characterized by the Chern numbers $(C_1, C_2, C_3)$.
    (b) Low-energy band structure at $g /\omega_{\mathrm{c}}=0.2$, and the band gap due to the cavity field at the $\mathrm{K}$ and $\Gamma $ points, respectively.
    (c)-(f) Low-energy band structures at $g /\omega_{\mathrm{c}}=1.65, 1.8, 2.02$ and $2.2$, respectively.
    (c) and (e) show the topological phase transitions at $g /\omega_{\mathrm{c}}=1.65$ and $2.02$, where the energy gap closes between the second and third bands at the $\Gamma$ point, and the first and second bands at the $\mathrm{K}$ point, respectively.
    The color of the bands indicates the expectation value of the photon number in (b)-(f).
    }
    \label{fig:phase_band_kagome_ED}
\end{figure*}

We investigate the band topology of the system with the muffin-tin potential
in the presence of the cavity field.
To reveal the band topology, we introduce a Chern number for each band, which is defined as follows~\cite{Nguyen2023_epChern}:
\begin{align}
    C_l = \frac{1}{2 \pi} \int_{\mathrm{BZ}} d^2 k \ i\big( \langle \partial_{k_x} &\Psi_{l \bm{k}} | \partial_{k_y} \Psi_{l \bm{k}} \rangle\nonumber \\
    &- \langle \partial_{k_y} \Psi_{l \bm{k}}| \partial_{k_x} \Psi_{l \bm{k}} \rangle \big),
    \label{eq:ChernNumber}
\end{align}
where $|\Psi_{l \bm{k}}\rangle$ is an electron-photon eigenstate for the $l$-th band.
We label bands in ascending order from the bottom unless a band touching occurs.
We characterize insulators with $l_{\mathrm{occ}}$ occupied bands
using the total Chern number $C$,
which is defined as the sum of the Chern numbers $C = \sum _{l=1}^{l_{\mathrm{occ}}}C_l$.
In this paper, 
we investigate the case where the low-energy band structures shown in Figs.~\ref{fig:periodic_bareband}(c) and ~\ref{fig:periodic_bareband}(d) are separated from other high-energy bands in the presence of light-matter interaction.
This condition is satisfied if the photon energy $\hbar \omega _{\mathrm{c}}$ is sufficiently larger than the bandwidth.
For example, choosing $\hbar \omega _{\mathrm{c}} = 0.2V_0$ fulfills this condition,
and the corresponding frequency is given by $\omega_{\mathrm{c}}=2\pi\times 494$ GHz for $V_0 = 10.2~\mathrm{meV}$.

In our parameter regime, we also compute the expectation value of the photon number $\langle \hat{a}^\dagger \hat{a}\rangle = \braket{\Psi _{l\bm{k}}|\hat{a}^\dagger \hat{a}|\Psi _{l\bm{k}}}$ for the low-energy bands (See Figs. \ref{fig:phase_band_kagome_ED} and \ref{fig:honeyMF_ED_band}).
The expectation value remains well below 1 for $g/\omega _{\mathrm{c}} \leq 3$, with a photon-number cutoff of 5.
Thus, we consider photon-number subspaces of up to 5 for diagonalizing the Hamiltonian in Eq.~(\ref{eq:QED_Hamil_Coulomb}),
which is sufficient to investigate the band topology.

\subsection{Kagome system case: $\alpha=1$}
We analyze the band topology in the kagome system by varying the coupling strength of the light-matter interaction.
We illustrate Chern numbers of the kagome systems in Fig.~\ref{fig:phase_band_kagome_ED} (a).
When the coupling strength is introduced,
the band degeneracy at the $\mathrm{K}$, $\mathrm{K}'$ and $\Gamma$ points is lifted [Fig.~\ref{fig:phase_band_kagome_ED} (b)].
Due to the mass gap, topologically nontrivial bands with $(C_1,C_2,C_3)=(1,0,-1)$ appear.

The light-matter interaction opens a band gap between the first and second bands.
Therefore, when the Fermi energy is inside the band gap between the first and second bands,
a Chern insulator with a total Chern number of $C=1$ is realized.
In contrast, when the light-matter interaction $g/\omega _{\mathrm{c}}$ is weak,
the third band, i.e., the flat band, can be topologically nontrivial with $C_3=-1$
while an indirect band overlap survives between the second and third bands.

Moreover, topological phase transitions between different Chern insulating phases occur in the USC regime. 
When $g / \omega_{\mathrm{c}} \simeq 1.65$,
the energy gap closes at the $\Gamma$ point, as shown in Fig.~\ref{fig:phase_band_kagome_ED} (c).
This gap closing gives rise to a topological phase transition that results in a topological band structure characterized by $(C_1, C_2, C_3)=(1, -2, 1)$.
Thus, this phase transition changes the sign of the Chern number of the topological flat band characterized by $C_3$.
Moreover, as shown in Fig.~\ref{fig:phase_band_kagome_ED} (d),
after the phase transition, the indirect band overlap between the second and third bands becomes smaller as the light-matter interaction increases.
Once the band gap entirely opens between the second and third bands,
the system is a Chern insulator with $C=C_1+C_2=-1$ if the Fermi energy lies within this gap.

Next, we assume that the Fermi energy lies in the band gap between the first and second bands.
When $g / \omega_{\mathrm{c}} \simeq 2.02$, the first and second bands touch at the $\mathrm{K}$ and $\mathrm{K}'$ points
[Fig.~\ref{fig:phase_band_kagome_ED} (e)],
which leads to topological bands with $(C_1, C_2, C_3)=(-1, 0, 1)$.
Thus, the system exhibits a Chern insulating phase with $C=C_1=-1$.
We show a band structure with $(C_1, C_2, C_3)=(-1, 0, 1)$ in Fig.~\ref{fig:phase_band_kagome_ED}(f).
As a result, the system undergoes topological phase transitions between the phases with $C = 1$ and $C = -1$ in the USC regime, when the Fermi energy lies in the lower band gap.
Therefore, we find that the band topology of the kagome systems can be changed by varying the coupling strength.
Even though we change parameters in the model,
such phase transitions can occur as long as the band gap above the third band remains open.

\subsection{Honeycomb system case: $\alpha=0$}
\begin{figure}[t]
    \centering
    \includegraphics[width=0.85\linewidth]{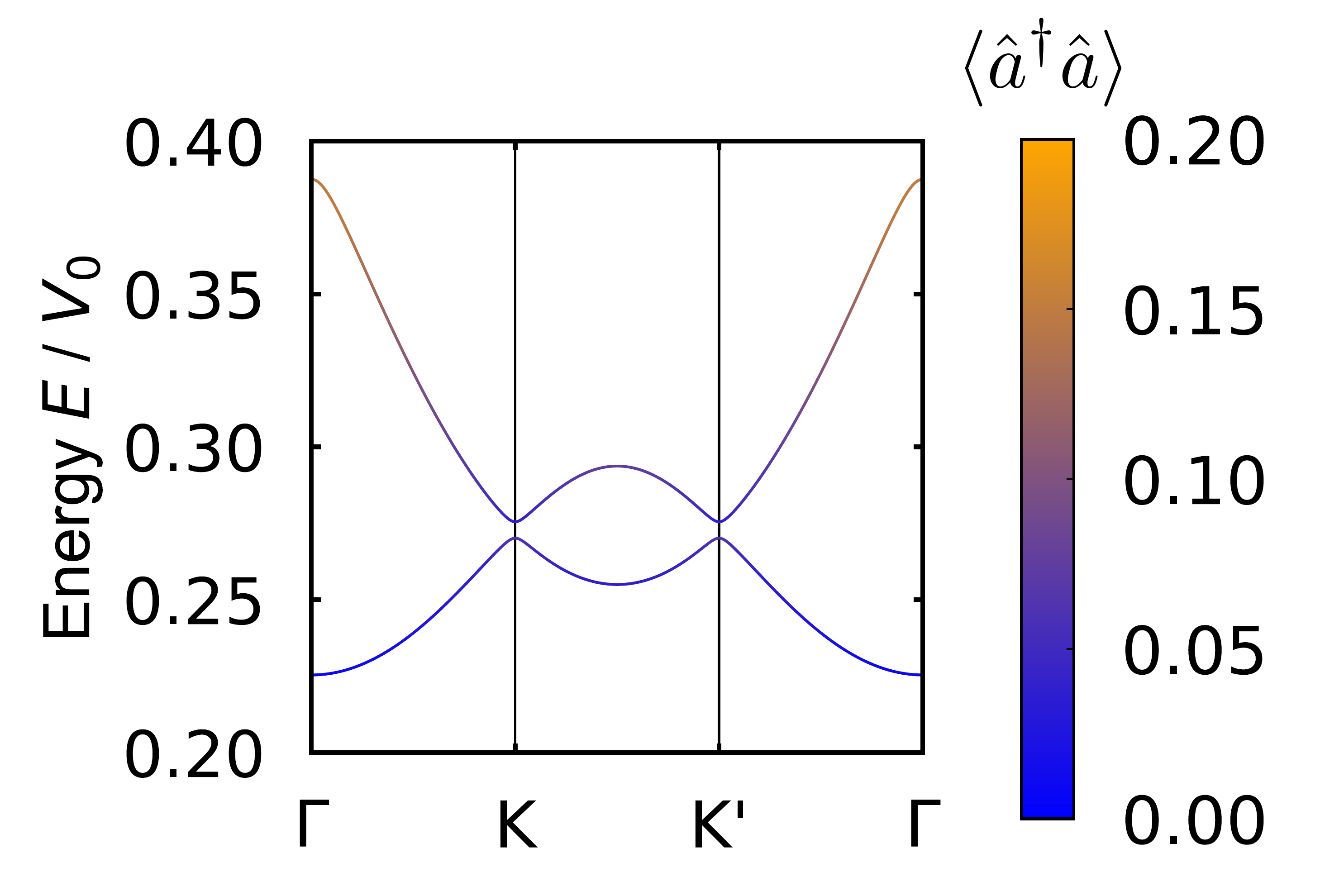}
    \caption{Low-energy band structure in the honeycomb system at the coupling strength $g /\omega_{\mathrm{c}}=0.5$.
    The color of the bands indicates the expectation value of the photon number.
    }
    \label{fig:honeyMF_ED_band}
\end{figure}

We next study band topology in the honeycomb system with the cavity field.
Figure~\ref{fig:honeyMF_ED_band} shows its band structure.
The light-matter coupling gives a mass gap to the Dirac points at the $\mathrm{K}$ and $\mathrm{K}'$ points,
which induces bands with $(C_1, C_2)=(1, -1)$. 
Thus, a Chern insulating phase with $C=1$ is realizable when only the lowest energy band is occupied.
However, unlike the kagome system,
no topological phase transition happens even though the light-matter coupling becomes stronger,
which is consistent with previous works \cite{Masuki2023_berryphase, Yang2025_EHM_topo}.
This behavior does not change, as long as the band gap above the second band remains open.

Here we compare the band topology of the cavity-embedded kagome and honeycomb systems in our parameter regime.
In the kagome system, the three-band structure enables gap closings between different pairs of bands as the coupling strength is increased.
Such gap closings allow Chern numbers $C_l$ to be redistributed among the three bands.
By contrast, in the honeycomb model, the gap closing occurs only between the two bands.
In view of the number of the bands,
this behavior is more likely in the kagome system than in the minimal two-band honeycomb model.
As a result, the cavity-embedded kagome system supports various topological bands.

\section{Topological edge states}
\label{sec:edge}

\begin{figure*}[t]
    \centering
    \includegraphics[width=1.0\linewidth]{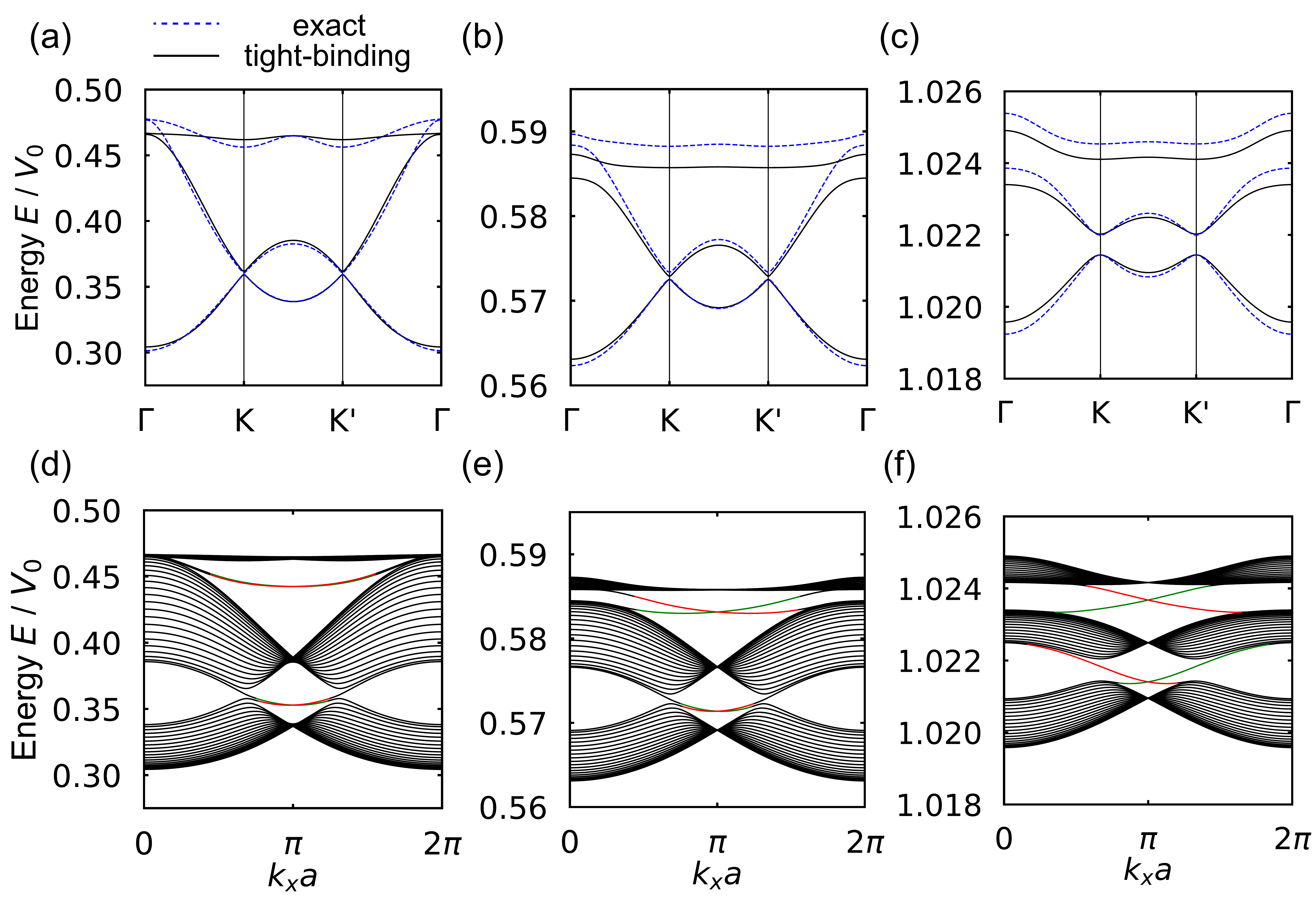}
    \caption{(a)-(c) Comparison of bulk band structures from the exact analysis and the tight-binding model at $g /\omega_{\mathrm{c}}=0.2, 1.8$, and $2.9$, which are topologically characterized by $(C_1, C_2, C_3)=(1,0,-1), (1,-2,1)$, and $(-1,0,1)$, respectively. 
    We adjust the lowest energy eigenvalue obtained from the tight-binding model at the $\mathrm{K}$ point to match that obtained from the exact analysis because the energy shift does not change the band topology.
    (d)-(f) Topological edge states for $g /\omega_{\mathrm{c}}=0.2, 1.8$, and $2.9$, respectively, with the periodic boundary condition in the $x$-direction and open boundary condition in the $y$-direction.
    Red and green solid lines represent gapless edge modes, and the color difference indicates that these modes are localized at opposite edges.
    }
    \label{fig:kagome_TB_band_edge}
\end{figure*}

Chern insulators exhibit topological edge states as a manifestation of the nontrivial bulk topology.
To see topological edge states, we construct a low-energy tight-binding model based on the cavity QED Hamiltonian in Eq.~\eqref{eq:QED_Hamil_Coulomb},
which is helpful to reduce computational cost.
First, we perform the asymptotically decoupling (AD) unitary transformation \cite{Ashida2021_ADframe, Masuki2023_berryphase},
which is given by
\begin{align}
    \hat{U}&=\exp \left( -i\xi _g \frac{\hat{\bm{p}}}{\hbar}\cdot \hat{\bm{\pi}} \right),
\end{align}
with
\begin{align}
    \hat{\bm{\pi}}=-i\bm{\varepsilon}\hat{a}+i\bm{\varepsilon}^*\hat{a}^{\dagger}
\end{align}
and
\begin{align}
    \xi_g &= \sqrt{\frac{\hbar}{m \omega_{\mathrm{c}}}} \frac{g / \omega_{\mathrm{c}}}{1 + g^2 / \omega_{\mathrm{c}}^2}.
    \label{eq:xi}
\end{align}
Then, the Hamiltonian in Eq.~(\ref{eq:QED_Hamil_Coulomb}) is transformed into
\begin{align}
\begin{split}
    &\hat{U}^{\dagger}\hat{H}_{\mathrm{C}}\hat{U} \\
    =&\frac{\hat{\bm{p}}^2}{2m_{\mathrm{eff}}}
    +V\left( \bm{r}+\xi _g\hat{\bm{\pi}} +\frac{\xi _g^2}{2\hbar}\hat{\bm{p}} \times \bm{e}_z \right ) 
    +\hbar \Omega \left( \hat{a}^{\dagger}\hat{a} + \frac{1}{2} \right) ,
    \end{split}
    \label{eq:UHU}
\end{align}
where $m_{\mathrm{eff}}=m(1+g^2/\omega _{\mathrm{c}}^2)$ is the effective mass dressed by cavity photons
and $\Omega = \omega _{\mathrm{c}} (1+g^2/\omega _{\mathrm{c}}^2)$ is the renormalized cavity frequency,
$\bm{e}_z$ is the unit vector in the $z$-direction.
In this AD frame, the effect of the quantum electromagnetic field is incorporated
into the electronic Hamiltonian through the effective mass and the modified periodic potential.
Since we focus on low-energy bands,
we project the AD-frame Hamiltonian onto the zero-photon subspace.
As a result, we obtain the effective low-energy Hamiltonian given by
\begin{align}
    \hat{H}^{\mathrm{AD}}_0 
    &=\braket{0|\hat{U}^{\dagger}\hat{H}_{\mathrm{C}}\hat{U}|0}
    \label{eq:UHU0}
\end{align}
where $\ket{0}$ is the photon vacuum.
This Hamiltonian allows us to construct a faithful tight-binding model in the USC regime \cite{Ashida2021_ADframe, Masuki2023_berryphase}.
In the weak-coupling limit $(g/\omega _{\mathrm{c}} \rightarrow 0)$ and in the strong-coupling limit $(g / \omega _{\mathrm{c}} \rightarrow \infty )$, $\xi _g$ vanishes.
Therefore, in the weak and strong coupling limits,
 $\hat{H}^{\mathrm{AD}}_0$ accurately reflects the topology of the low-energy bands
as long as the system preserves energy gaps between adjacent bands with different photon numbers \cite{Masuki2023_berryphase}.
The details of this approach are shown in Appendix \ref{ap:AD}.

By obtaining Wannier functions from the low-energy Hamiltonian $\hat{H}^{\mathrm{AD}}_0$, we can construct a low-energy tight-binding model for the kagome and honeycomb systems.
Using Wannier90, we compute maximally localized Wannier functions
\cite{Marzari1997_MLWF, Souza2001_MLWF, Marzari2012_MLWF, Pizzi2020_MLWF}
and obtain hopping amplitudes for the tight-binding models. 
In the following, we demonstrate the existence of the topological edge states using the tight-binding model for the kagome system.
We also show edge states in the honeycomb Chern insulator in Appendix \ref{ap:graphene}.

Using this approach, we construct a three-band tight-binding model for the kagome system with the cavity field.
Here, we consider up to the third nearest neighbor hopping,
and calculate the bulk band structure. 
Figures~\ref{fig:kagome_TB_band_edge} (a)-(c) show the comparison between the energy bands obtained from the exact analysis and the tight-binding model.
The tight-binding model can accurately reproduce the two low-energy bands calculated in the exact analysis.
The third band obtained from the tight-binding model differs slightly because high-energy photon subspaces are ignored.
Nevertheless, the tight-binding model exhibits the same topological band structures characterized
by $(C_1, C_2, C_3)=(1, 0, -1), (1, -2, 1)$ and $(-1, 0, 1)$, as shown in Fig.~\ref{fig:phase_band_kagome_ED} (a).
We show details of band topology of the tight-binding model in Appendix \ref{ap:AD}.

Since the tight-binding model can describe the band topology,
we calculate energy bands with an open boundary condition to see topological edge states.
The number of topological edge modes is determined by the total Chern number.
As shown in Figs.~\ref{fig:kagome_TB_band_edge} (d)-(f),
we can find topological edge states between the bulk gaps.
We note that topological edge modes can appear in the upper band gap even though an indirect band overlap exists between the second and third bulk bands.
The emergence of these edge states reflects the band topology of the first and second bands characterized by $C_1+C_2$.
At the same time, the sign of the total Chern number also specifies their chirality, that is, the direction of the edge current.
Therefore, the chirality of the edge states is switched by changing the light-matter coupling.
The topological edge modes change the chirality at the topological phase transition because of the gap closing.

\section{Conclusion and Outlook}
\label{sec:conc}

We have studied the band topology of a kagome system using the muffin-tin potential in a chiral cavity.
Because the cavity field, which breaks time-reversal symmetry, induces a mass gap,
a Chern insulating phase emerges when the light-matter interaction is introduced.
We have shown that a flat band acquires a nonzero Chern number due to the light-matter interaction.
Moreover, we have found that stronger light-matter coupling can induce topological phase transitions between different Chern insulating phases in the kagome system.
These changes in the band topology stem from gap closings between different pairs of the three bands, suggesting that the presence of multiple bands is important.
We also have demonstrated existence of topological edge states by constructing tight-binding models.
The chirality of the edge states is switched when the topological phase transition occurs in the USC regime.
Our work has revealed that various Chern insulating phases can be realized in the cavity-embedded kagome system.
Experimentally, quantum Hall transport has been measured in split-ring resonator cavities \cite{Appugliese2022_breakdown_topo, Enkner2025_cavity_qhe}.
Therefore, the kagome Chern insulators induced by cavity fields can be observed via quantum Hall transport.

We have investigated topological band structures in a cavity-embedded kagome system.
For simplicity, 
we have assumed that three low-energy bands with a narrow bandwidth are isolated from other energy bands in the presence of light-matter interaction.
Recently, kagome-like band dispersions with a narrow bandwidth have been experimentally engineered in 2DEGs with patterned electrostatic potentials \cite{wang2024tuning}.
Moreover, chiral cavities have been experimentally realized \cite{Andberger2024, Aupiais2024,DanielSciAdv,kulkarni2025realization}. 
Therefore, in the (sub-)THz regime, 2DEGs embedded in chiral cavities can be regarded as a promising platform for realizing such cavity-induced topological phases
\footnote{
The vector potential amplitude $A_0$ is estimated as $A_0=\sqrt{\hbar /(2\epsilon _0\epsilon _\mathrm{r} \omega _{\mathrm{c}} V_{\mathrm{eff}})}$, where $\epsilon _0$ is the vacuum permittivity, $\epsilon _\mathrm{r}$ is the relative permittivity, and $V_{\mathrm{eff}}$ is the effective cavity mode volume.
For InSb with $m \sim 0.01 m_\mathrm{e}$ and $\epsilon_{\mathrm{r}} = 15.7$,
we can obtain the estimate $g/\omega _{\mathrm{c}} \sim 0.1$ for $V_{\mathrm{eff}} \sim 10^{-18}\ \mathrm{m}^3$ in the sub-THz regime.
This suggests that the USC regime may be approached in cavities with a small effective mode volume, such as a metamaterial cavity with dipole nanoantennas \cite{Andberger2024} and the Faraday-rotator-based cavity \cite{Hubener2021}.
}.

For realistic materials, 
the above assumption is generally not satisfied in typical kagome materials
\footnote{
To satisfy the conditions in realistic materials with lattice constant $a \simeq 5~$\AA ~in the THz regime, an effective mass ratio $m /m_\mathrm{e}$ of order $10^2$ is required in our model calculations. 
},
although narrow bandwidths may be realized in heavy-fermion kagome materials \cite{Song2025,lee2025arXiv}.
Nevertheless, we expect that typical kagome systems with Dirac points
exhibit topological bands with nonzero Chern numbers in a chiral cavity
since the light-matter interaction opens a band gap due to time-reversal symmetry breaking.
Indeed, Dirac points have been experimentally observed in kagome materials, such as CsV$_3$Sb$_5$ \cite{Ortiz2020_kagome, Liu2021_kagome, Lou2022_kagome}.
Moreover, realistic kagome materials possess multiple bands near the Dirac points and flat band.
Based on our results, the presence of multiple bands is expected to enrich the band topology through gap closings between different band pairs.

Finally, we comment on several effects that may render topological phase transitions accessible at moderate coupling strength.
In realistic kagome materials,
the photon vacuum can strongly hybridize with higher-energy $n$-photon sectors ($n>0$) in the (sub-)THz regime.
This hybridization may induce topological phase transitions that lead to various topological band structures beyond those considered in this paper.
Another possible route is to include multiple cavity modes.
Because the effective mass $m_{\mathrm{eff}}$ can be enhanced by multiple modes \cite{Masuki2023_berryphase, Jiang2025_cavity},
this multimode effect can modify the band structures and may enable different topological phase transitions.
A detailed analysis of these hybridization and multimode effects is left for future work.

\begin{acknowledgments}
H.G. is grateful to K. Masuki, K. Saito and T. Nakamoto for fruitful discussions.
This work was supported by JSPS KAKENHI (Grant No. JP23K13033, No. JP24K00586, and No. JP25H01248).
\end{acknowledgments}

\appendix
\section{Low-energy Hamiltonian and band topology in the AD frame}
\label{ap:AD}

\subsection{Construction of the low-energy Hamiltonian}
Before detailing the construction of the low-energy Hamiltonian in the kagome and honeycomb systems,
we briefly explain the reason for using the AD frame \cite{Ashida2021_ADframe, Masuki2023_berryphase}.
When $g/\omega _{\mathrm{c}} \gtrsim 0.1$, the Peierls substitution may not accurately produce energy bands of the cavity QED Hamiltonian \cite{Ashida2021_ADframe, Masuki2023_berryphase}.
Therefore, in such an USC regime,
we need a different approach to construct a low-energy tight-binding model.
In the AD frame,
$\xi_g$ vanishes both in the weak-coupling limit $(g/\omega _{\mathrm{c}} \rightarrow 0)$ and in the strong-coupling limit $(g / \omega _{\mathrm{c}} \rightarrow \infty)$, 
as seen in Eq.~(\ref{eq:xi}).
Because electronic and photonic degrees of freedom are asymptotically decoupled in these limits, 
low-energy eigenstates can be approximated by the product state of an electronic wavefunction and the photon vacuum.
Hence, the effective Hamiltonian in Eq.~(\ref{eq:UHU0}) can reproduce the low-energy band structure in the strong-coupling limit, which is useful to construct the tight-binding model.

We here describe the detailed construction of the low-energy Hamiltonian in the kagome and honeycomb systems in the AD frame. 
By applying the Fourier transformation to the potential, the Hamiltonian in Eq.~(\ref{eq:UHU0}) becomes
\begin{align}
    \hat{H}^{\mathrm{AD}}_0
    =\frac{\hat{\bm{p}}^2}{2m_{\mathrm{eff}}}
    +\sum _{\bm{G}}V_{\bm{G}}e^{-\frac{\xi _g^2G^2}{4}}
    e^{i\bm{G}\cdot (\bm{r} + \frac{\xi _g^2}{2\hbar }\cdot \hat{\bm{p}}\times \bm{e}_z)}
    +\frac{1}{2}\hbar \Omega,
\end{align} 
where $\bm{G} = m_1\bm{b}_1+m_2\bm{b}_2=(G_x, G_y)$ is a reciprocal lattice vector
with integers $m_1$ and $m_2$ and $\bm{b}_1=2\pi /a(1, 1/\sqrt{3})$ and $\bm{b}_2=2\pi /a(0,2/\sqrt{3})$. 
Here, 
\begin{align}
    V_{\bm{G}}=\frac{1}{S_{\mathrm{cell}}}
    \int _{\mathrm{cell}}d^2r V(\bm{r})e^{-i\bm{G} \cdot \bm{r}},
\end{align}
where $S_{\mathrm{cell}}=\sqrt{3}a^2/2$ is the area of the unit cell,
and the integration is performed over the unit cell.
For the potential in Eq.~(\ref{eq:muffin-tin_potential}), $V_{\bm{G}}$ is given by \cite{Xu2024_kagome_Muffintin}
\begin{align}
    V_{\bm{G}} = \frac{2\pi V_0}{S_{\mathrm{cell}} |\bm{G}|} [ \rho_{\mathrm{H}} &J_1\left( \rho_{\mathrm{H}} |\bm{G}| \right)\nonumber\\
     &+ 2\alpha \rho_{\mathrm{T}} J_1\left( \rho_{\mathrm{T}} |\bm{G}| \right) \cos\left( \bm{G}\cdot\bm{L} \right) ],
    \label{eq:Fourier_PP_kagome}
\end{align}
where $J_1$ is the Bessel function of the first kind.
In momentum space, the matrix elements of the Hamiltonian $\hat{H}^{\mathrm{AD}}_0(\bm{k})=e ^{-i\bm{k}\cdot\bm{r}} \hat{H}^{\mathrm{AD}}_0 e^{i\bm{k}\cdot\bm{r}}$ are represented by
\begin{widetext}
    \begin{align}
\braket{\bm{\mathcal{K}}|\hat{H}_0^{\mathrm{AD}}(\bm{k})|\bm{\mathcal{K}}'}
=\left[ \frac{\hbar^2 \left( \bm{k} + \bm{\mathcal{K}} \right)^2}{2 m_{\mathrm{eff}}}
 +\frac{1}{2}\hbar\Omega \right] \delta_{\bm{\mathcal{K}},\bm{\mathcal{K}}'}
+\sum _{\bm{G}}V_{\bm{G}}e^{-\frac{\xi _g^2\bm{G}^2}{4}}e^{i\frac{\xi _g^2}{2}[\bm{G}\times (\bm{k} + \bm{\mathcal{K}})]_z}
\delta_{\bm{\mathcal{K}}, \bm{\mathcal{K}}' + \bm{G}},
\end{align}
\end{widetext}
where $\bm{\mathcal{K}}=n_1\bm{b}_1+n_2\bm{b}_2$ is a reciprocal lattice vector with integers $n_1$ and $n_2$,
and $\ket{\bm{\mathcal{K}}}$ is the plane wave state with wavevector $\bm{\mathcal{K}}$.
By diagonalizing the Hamiltonian in the zero-photon subspace, we can obtain low-energy electronic bands to compute Wannier functions.
Using these results, we can construct tight-binding models of the kagome and honeycomb systems.

\begin{figure}[b]
    \centering
    \includegraphics[width=1.0\linewidth]{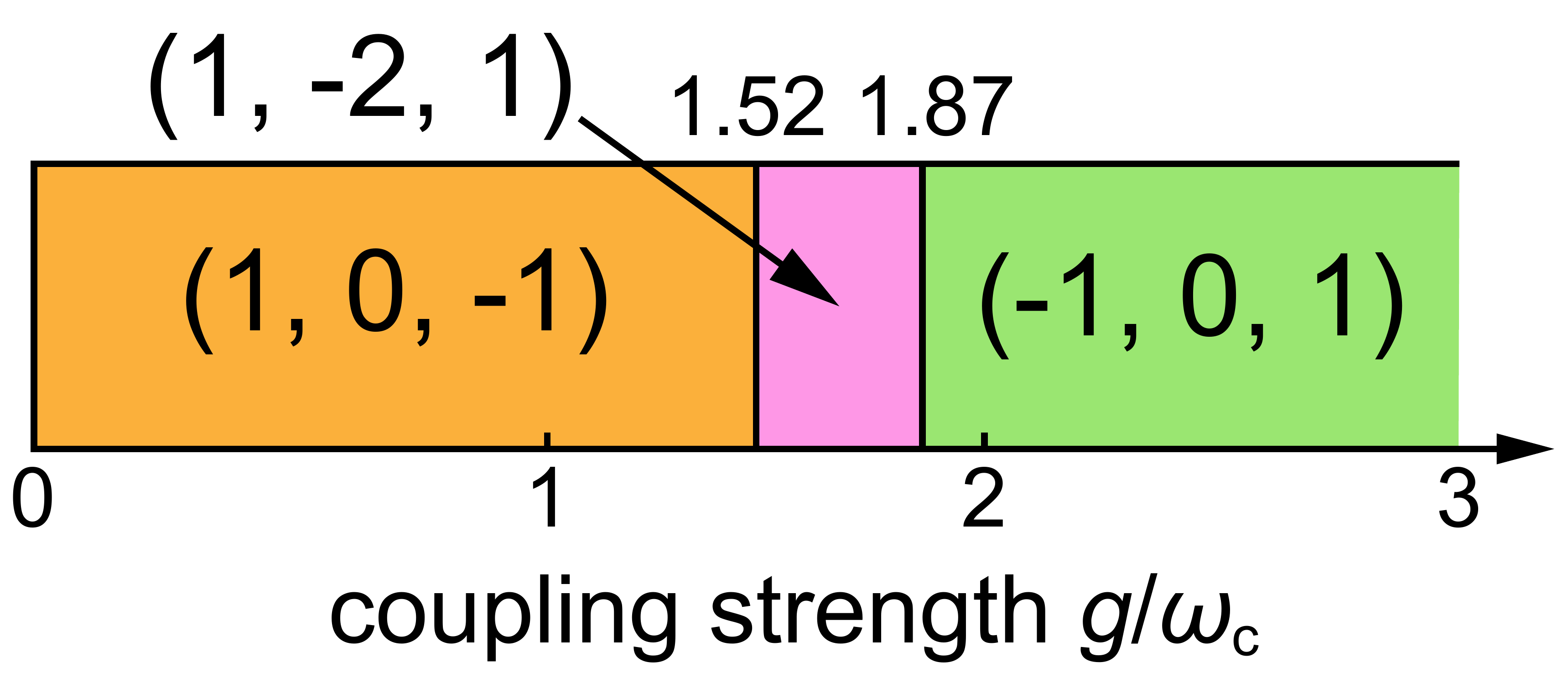}
    \caption{Chern numbers of the tight-binding model for the kagome system.
    }
    \label{fig:kagome_TB_PhaseDiagram}
\end{figure}

\subsection{Band topology of the tight-binding model}
In addition, we discuss the band topology of the tight-binding model for our system over a broad range of coupling strengths.
We present Chern numbers obtained from the tight-binding model for the kagome system in Fig.~\ref{fig:kagome_TB_PhaseDiagram}.
This band topology agrees with that in Fig.~\ref{fig:phase_band_kagome_ED} (a)
although we do not consider the hybridization between the photon vacuum $\ket{0}$ and the $n$-photon states $\ket{n}$ for $n>0$.
In the AD frame, the matrix element of $\hat{H}^{\mathrm{AD}}:=\hat{U}^{\dagger}\hat{H}_{\mathrm{C}}\hat{U}$ between $\ket{0}$ and $\ket{n}$ for $n>0$ is given by
\begin{align}
&\braket{\bm{\mathcal{K}}\otimes 0|\hat{H}^{\mathrm{AD}}(\bm{k})|\bm{\mathcal{K}}'\otimes n} \notag \\
&=\sum _{\bm{G}}V_{\bm{G}}e^{i\frac{\xi _g^2}{2}[\bm{G}\times (\bm{k}+\bm{\mathcal{K}})]_z}\langle 0|e^{i\xi_g\bm{G}\cdot\hat{\bm{\pi}}}|n\rangle
\delta_{\bm{\mathcal{K}}, \bm{\mathcal{K}}' + \bm{G}},
\end{align}
where $\ket{\bm{\mathcal{K}}\otimes 0}$ is a product state of
the plane wave state with wavevector $\bm{\mathcal{K}}$ and the photon
vacuum, and
\begin{align}
    \langle 0|e^{i\xi_g\bm{G}\cdot\hat{\bm{\pi}}}|n\rangle
    =\frac{1}{\sqrt{n!}}\left[ \frac{\xi_g (G_x-iG_y)}{\sqrt{2}}\right] ^n
    e^{-\frac{\xi_g^2\bm{G}^2}{4}}.
\end{align}
Thus, when $\xi _g/a$ is vanishingly small,
the hybridization between the photon vacuum and the $n$-photon sectors becomes negligible,
 and does not significantly affect the band topology.
According to Eq.~(\ref{eq:xi}), the effective coupling strength decreases as the frequency $\omega _{\mathrm{c}}$ is higher.
In our frequency regime, this hybridization does not change the possible topological phases.  
Therefore, because $\hat{H}^{\mathrm{AD}}_0$ can describe the low-energy band topology of our kagome system,
the tight-binding model can reproduce its topological phases.

\begin{figure}[b]
    \centering
    \includegraphics[width=1.0\linewidth]{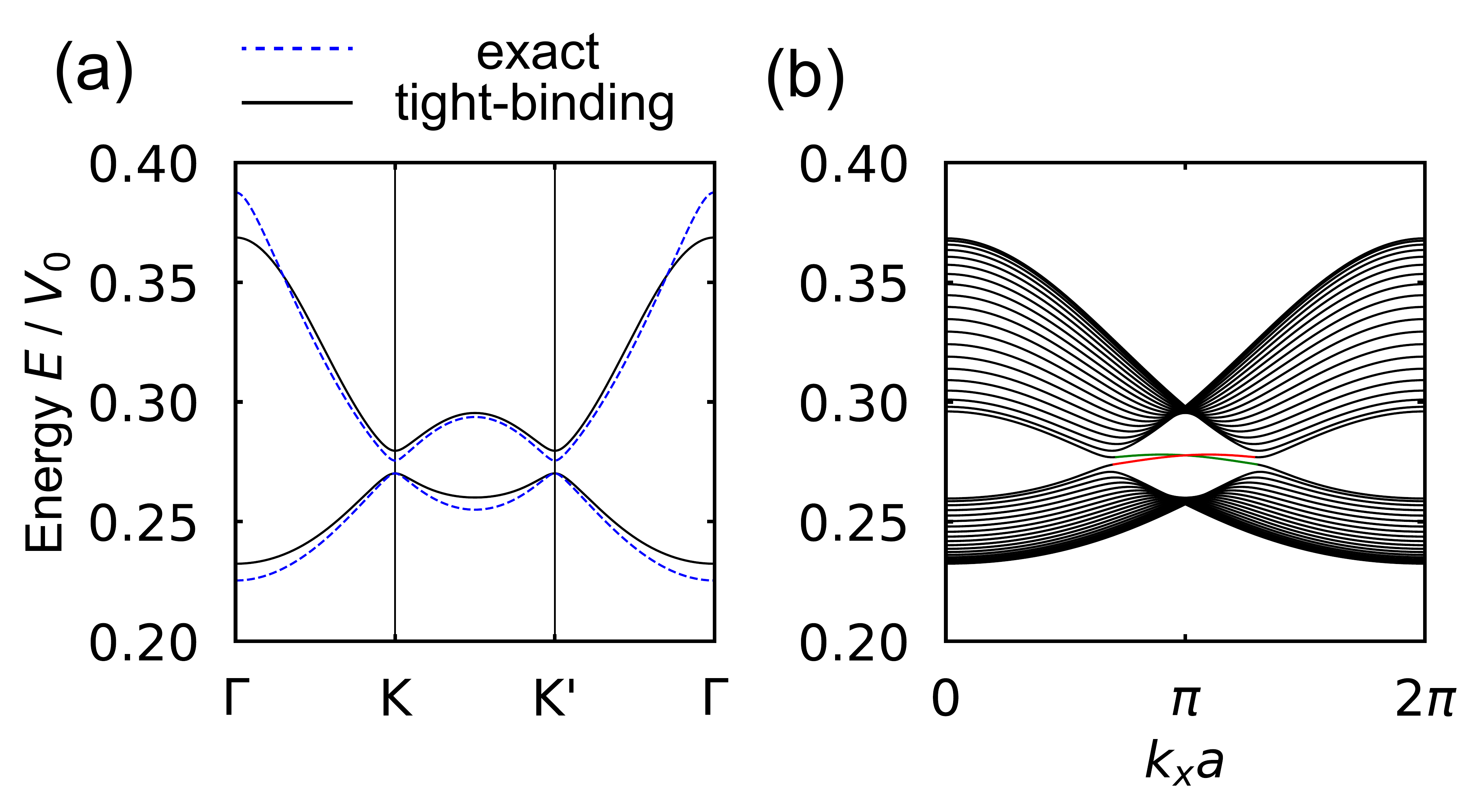}
    \caption{(a) Band structure in the tight-binding model for the honeycomb system at $g /\omega_{\mathrm{c}}=0.5$.
    We shift the lowest energy eigenvalue at the $\mathrm{K}$ point to match that obtained from the exact analysis.
    (b) Topological edge states with the periodic boundary condition in the $x$-direction and open boundary condition in the $y$-direction.
    Red and green solid lines represent gapless edge modes, and the color difference indicates that these modes are localized at opposite edges.}
    \label{fig:honeyMF_TB_band_edge}
\end{figure}

\section{Edge states in the honeycomb Chern insulator}\label{ap:graphene}
We construct a low-energy tight-binding model for the honeycomb system.
In the same way as the kagome system,
we include hopping up to the third nearest neighbors.
The bulk band structure is shown in Fig.~\ref{fig:honeyMF_TB_band_edge} (a).
The tight-binding Hamiltonian exhibits the same Chern insulating phase as that in the exact analysis,
which is characterized by $(C_1, C_2)=(1,-1)$.
Figure~\ref{fig:honeyMF_TB_band_edge} (b) shows topological edge state calculated from the tight-binding model.
Therefore, the bulk-edge correspondence is confirmed in the honeycomb lattice.
However, the chirality of the edge states is not tunable by the light-matter coupling in the high-frequency regime.


\bibliography{cavity}
\end{document}